\documentclass[journal]{IEEEtran}
\pdfoutput=1 
\ifCLASSINFOpdf
\else
   \usepackage[dvips]{graphicx}
\fi
\usepackage{url}
\usepackage{graphicx}
\usepackage[utf8]{inputenc} 
\usepackage[T1]{fontenc}
\usepackage{url}
\usepackage{bm}
\usepackage[cmex10]{amsmath} 
\usepackage{ifthen}
\usepackage{cite}
\usepackage{dsfont}
\usepackage{color}
\usepackage{graphicx}
\usepackage{color,soul}
\usepackage{amssymb}
\usepackage{amsthm}
\usepackage{amsfonts}
\usepackage{comment}
\usepackage[utf8]{inputenc} 
\usepackage[T1]{fontenc}
\usepackage{url}
\usepackage{ifthen}
\usepackage{cite}
\usepackage{color}
\usepackage{graphicx}
\usepackage{amssymb}
\usepackage{amsthm}
\usepackage{amsfonts}
\usepackage{comment}
\usepackage{subfig}

\newtheorem{theorem}{Theorem}
\usepackage[skip=5pt,style=base]{caption}

\usepackage{bm}
\usepackage[cmex10]{amsmath} 

\DeclareMathOperator{\E}{\mathbb{E}}
\DeclareMathOperator{\C}{\mathbb{C}}

\usepackage[skip=5pt,style=base]{caption}
\newtheorem{remark}{Remark}

\newtheorem{assumption}{Assumption}

\begin{document}
\title{Efficient model selection in switching linear dynamic systems by graph clustering}


\author{Parisa Karimi,  Mark D.\ Butala, Zhizhen Zhao, and Farzad Kamalabadi
\thanks{P.\ Karimi$^*$, Z.\ Zhao, and F.\ Kamalabadi are with the University of Illinois at Urbana-Champaign, Urbana, IL. 61801, USA (e-mail$^*$: parisa2@illinois.edu).}
\thanks{M.\ Butala is with the Zhejiang University, Hangzhou, Zhejiang, P.\ R.\ China.}}

\maketitle
\begin{abstract}
The computation required for a switching Kalman Filter (SKF) increases exponentially with the number of system operation modes. In this paper, a computationally tractable graph representation is proposed for a switching linear dynamic system (SLDS) along with the solution of a minimum-sum optimization problem for clustering to reduce the switching mode cardinality offline, before collecting measurements. It is shown that upon perfect mode detection, the induced error caused by mode clustering can be quantified exactly in terms of the dissimilarity measures in the proposed graph structure. Numerical results verify that clustering based on the proposed framework effectively reduces model complexity given uncertain mode detection and that the induced error can be well approximated if the underlying assumptions are satisfied.
\end{abstract}
\begin{IEEEkeywords}
Dynamic systems, Kalman filter, Recursive estimation, Model mismatch, Model complexity, 
Graph clustering.
\end{IEEEkeywords}

\IEEEpeerreviewmaketitle

\vspace{-.5cm}

\section{Introduction}


State estimation of a dynamical system from a collection of indirect, noisy measurements and a dynamic evolution model is a problem found in virtually all branches of engineering and physical sciences, e.g., \cite{rem1,rem2,power,geo,gene}. Diverse applications ranging from, e.g., financial forecasting, multiple target tracking, and neural pattern discrimination \cite{fin,mtt,eeg} are faithfully represented as a switching linear dynamic system (SLDS) obtained by augmenting hidden discrete random variables to a linear state space model \cite{dbn}. Joint estimation and detection of the hidden system mode can be accomplished with a switching Kalman filter (SKF) \cite{skf,skf2,skf3}.


The computational burden of a SKF increases exponentially with the number of system modes, hence rendering its utility impractical when the number of dynamical modes and the state variable dimension are large. In this work, we reduce the number of system modes by combining those that minimize the ultimate impact on the estimation error. Our approach involves the systematic modeling of an SLDS as a fully connected graph based on information determined using online/batch system identification algorithms\cite{si_alg,si-bay,hdp,sis-on1}. Node clustering, informed by the estimation error tradeoff, is then used to reduce the system mode cardinality before collecting the measurements and purely based on the model and the priors.

Our approach relates to efficient mode selection for a Gaussian mixture (GM) distribution, a task for which many algorithms have been developed to learn the GM parameters from data using e.g., expectation-maximization (EM), and for learning model complexity via the Bayesian information criterion (BIC) or Akaike information criterion (AIC)\cite{bishop}. Other approaches range from heuristic measure-based clustering\cite{gmm1,wd} and covariance union methods \cite{gmm2} to clustering using K-L divergence based measures\cite{gmm3}. However, our work in this paper addresses the more challenging problem of GM mode selection in dynamic scenarios (spatial and temporal) as opposed to the static (spatial-only) context. This work proposes an efficient algorithm to reduce the number of modes of an SLDS before observing the measurements, which would allow the user to run an SLDS with fewer modes in operation.

The remainder of the paper is organized as follows. The SLDS signal model and KF/SKF formulations are reviewed in Sections~\ref{section:slds} and~\ref{section:skf}, respectively. Section~\ref{section:graph} introduces the proposed graph structure as well as a dissimilarity measure for model complexity reduction. The effect of clustering on computational complexity is given in Section~\ref{Section:comp}. Simulations investigate the algorithm performance in Section~\ref{section:sim} and conclusions are given in Section~\ref{section:con}. 


\vspace{-.7cm}
\section{Signal model}\label{section:slds}

We adopt the following notational conventions:\\
\textbullet~$\bm{x}\sim \mathcal{N}(\bm{m},\bm{C})$: the random vector $\bm{x}$ has a Gaussian distribution with mean $\bm{m}$ and covariance $\bm{C}$.\\
\textbullet~$\E,\C,p$ refer to expectation, covariance, and probability, respectively, $\bm{I}$ refers to the identity matrix, and superscript $T$ refers to matrix transposition.\\
\textbullet~$\mathds{1}(S)=1$ if the statement $S$ is true, and it is zero otherwise.
The state-space model for an SLDS is
\begin{align}
    \bm{x}_n &= \bm{A}_n \bm{x}_{n-1}+\bm{\nu}_n, \label{eqn:sp1}\\
    \bm{y}_n &= \bm{H}_n\bm{x}_n +\bm{\omega}_n \label{eqn:sp2},
\end{align}
where $\bm{\nu}_n \sim \mathcal{N}(\bm{0},\bm{Q}_n)$ and $\bm{\omega}_n \sim \mathcal{N}(\bm{0},\bm{R}_n)$ are additive Gaussian noise such that $\bm{\nu}_n \perp \bm{\nu}_{n'}\,\forall n\neq n'$, $\bm{\omega}_n \perp \bm{\omega}_{n'}\,\forall n\neq n'$, $\bm{\nu}_n \perp \bm{\omega}_{n'}\, \forall n,n'$, and  $\bm{Q}_n=\E[\bm{\nu}_n\bm{\nu}_{n}^T]$, $\bm{R}_n=\E[\bm{\omega}_n\bm{\omega}_{n}^T]$.  The evolution matrices $\bm{A}_n= \bm{A}({s_n})$ are randomly drawn from the set of $z\times z$ evolution matrices $\mathcal{A}=\{ \bm{A}^{b}:b=1,\dots,r\}$, the covariance matrices $\bm{Q}_n = \bm{Q}({s_n})$ are randomly drawn from the set $\mathcal {Q}= \{\bm{Q}^{b}:b=1,\dots,r\}$, and the random variable $s_n \in \{ 1, \dots, r\}$ refers to the mode of the dynamic system at time $n$, which is drawn from a discrete Markov process assuming known initial prior and transition matrix. The hidden $z$-dimensional state variable at time step $n$, i.e., $\bm{x}_n \in \mathbb{R}^z$, is then estimated given the $q$-dimensional measurement vector $\bm{y}_n$, the $q\times z$ measurement matrix $\bm{H}_n$, $\mathcal {Q}, \mathcal {A}$, and $\bm{R}_n$. ($q$ is the number of measurements, $z$ is the state dimension, and $r$ is the number of modes the system may switch between).

%

\vspace{-12pt}
\section{State Estimators}\label{section:skf}
 Let $\bm{y}_1^n\equiv \{\bm{y}_1,\bm{y}_2,...,\bm{y}_n\}$, $\bm{x}_0\sim \mathcal{N}(\bm{x}_{0|0},\bm{P}_{0|0})$, $\bm{x}_{n|n}=\E[\bm{x}_n|\bm{y}_1^n]$, and $\bm{P}_{n|n}=\C(\bm{x}_n|\bm{y}_1^n)$. The $Filter$ operator is
\begin{multline}
    (\bm{x}_{n|n},\bm{P}_{n|n}) = Filter(\bm{A}_n,\bm{H}_n,\bm{x}_{n-1|n-1}, \\\bm{P}_{n-1|n-1}, \bm{Q}_n,\bm{R}_n,\bm{y}_1^n) \label{eqn:filter}
\end{multline}
which involves the repeated application of a time update step
\begin{align}
\bm{x}_{n|n-1} &= \bm{A}_n \bm{x}_{n-1|n-1} \label{eqn: Kalman1}\\
\bm{P}_{n|n-1}&= \bm{A}_n \bm{P}_{n-1|n-1}\bm{A}_n^T + \bm{Q}_n
\end{align}
and a measurement update step
\begin{align}
\bm{K}_n &= \bm{P}_{n|n-1} \bm{H}_n^T \bigl(\bm{H}_n \bm{P}_{n|n-1} \bm{H}_n^T + \bm{R}_n\bigr)^{-1}\\
\bm{x}_{n|n} &= \bm{x}_{n|n-1} + \bm{K}_n \bigl(\bm{y}_n - \bm{H}_n \bm{x}_{n|n-1}\bigr)\\
\bm{P}_{n|n} &= (\bm{I} - \bm{K}_n \bm{H}_n) \bm{P}_{n|n-1}. \label{eqn:Kalmanend} 
\end{align}
  The application of a single-mode KF to the general SLDS \eqref{eqn:sp1}-\eqref{eqn:sp2} results in erroneous estimates. The well known SKF detects the switching mode at each time step and its corresponding model parameters ($\bm{A}_n,\bm{Q}_n$) using a likelihood ratio test at each time step, and estimates the state variables accordingly. Upon perfect detection of the modes, one can obtain optimal estimates of the state variable $\bm{x}_n$ in terms of both maximum a posteriori (MAP) and mean squared error (MSE) metrics; however, due to the exponentially increasing computational cost of exact SKFs, approximate SKFs \cite{skf} based on a finite memory of past trajectories are used in practice to provide suboptimal but feasible estimates. The SKF algorithm investigated throughout this paper when considering a multi-modal dynamic model refers to the MAP estimator, which implements the KF corresponding to the mode with the largest likelihood at each time step. 

 \begin{figure}
  \centering
    \includegraphics[width=0.4\textwidth]{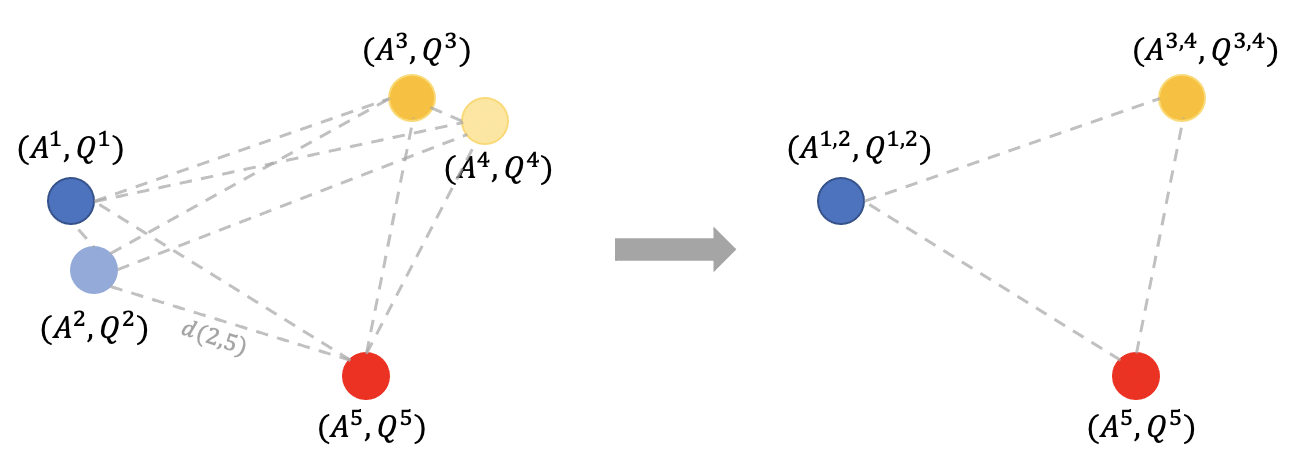}
    \caption{ Using the proposed graph, $5$-modal SLDS model with evolution matrix and covariance matrix pair $(A^i,Q^i),\forall i\in\{1,\dots,5\}$ may be reduced to a $3$-modal SLDS model, where $(A^{i,j},Q^{i,j})$ refers to the evolution and covariance matrix pair of the mode obtained by merging modes $i,j$. }
    \label{fig:rmodal}
        \vspace{-12pt}
\end{figure} 

\vspace{-7pt}
\section{Graph representation of the SLDS}\label{section:graph}
An SLDS may be fully represented by a graph based on its switching mode parameters, initial values, and detection rates. At a given time step $n$, the SLDS graph representation is as follows. The nodes are the current linear dynamic system switching parameters, represented by $(\bm{A}^{i},\bm{Q}^{i}), i=1,2,...,r$, as well as their prior probabilities and the transition probabilities at each time step. 
%
Let $\{S\}$ denote the set of modes used for SKF estimation when the true underlying modes are $\{S'\}$. Then, the MSE is $\epsilon_{\{S\}|\{S'\}}^{(n)} = \E\bigl[(\bm{x}_n-\hat{\bm{x}}_{n|n})^T (\bm{x}_n-\hat{\bm{x}}_{n|n})\mid \{S\}, \{S'\} \bigr]$, where $\bm{x}_n$, $\hat{\bm{x}}_{n|n}$ refer to the ground truth and estimated state variable at time $n$, respectively.
 
 The directed edge dissimilarity measure from node $j$ to $i$, $d^{(n)}(j\rightarrow i)={ \epsilon_{\{i\}|\{i,j\}}^{(n)}- \epsilon^{(n)}_{\{i,j\}|\{i,j\}}}$, is defined as the additional estimation error incurred in a bimodal system with modes $i,j$ when using statistics from a single node $\{i\}:\{(\bm{A}^{i},\bm{Q}^{i})\}$ instead of both nodes $\{i,j\}:\big\{(\bm{A}^{i},\bm{Q}^{i}),(\bm{A}^{j},\bm{Q}^{j})\big\}$. When the likelihood of the true mode is maximized, it can be verified that ${ \epsilon^{(n)}_{\{i\}|\{i,j\}}- \epsilon^{(n)}_{\{i,j\}|\{i,j\}}} = \frac{\pi_j}{\pi_i+\pi_j}({ \epsilon^{(n)}_{\{i\}|\{j\}}- \epsilon^{(n)}_{\{j\}|\{j\}}})$. Thus, the undirected edge dissimilarity measure $d^{(n)}(i,j)$ between nodes $i,j$ may be defined: 
\begin{align}
     &d^{(n)}{(i,j)} = d^{(n)}(i\rightarrow j)+d^{(n)}(j\rightarrow i)\nonumber \\
     &=\bigl({ \epsilon^{(n)}_{\{j\}|\{i,j\}}- \epsilon^{(n)}_{\{i,j\}|\{i,j\}}}\bigr)+ \bigl({ \epsilon^{(n)}_{\{i\}|\{i,j\}}- \epsilon^{(n)}_{\{i,j\}|\{i,j\}}}\bigr)\nonumber \\
     &=\frac{\pi_i}{\pi_i+\pi_j}\bigl({ \epsilon^{(n)}_{\{j\}|\{i\}}- \epsilon^{(n)}_{\{i\}|\{i\}}}\bigr)+ \frac{\pi_j}{\pi_i+\pi_j}\bigl({ \epsilon^{(n)}_{\{i\}|\{j\}}- \epsilon^{(n)}_{\{j\}|\{j\}}}\bigr), \label{eqn:edge}
 \end{align}
 where $\pi_i$ is the occurrence probability of mode $i$. The justification for this measure is provided in Theorem~\ref{theorem:1}/Remark~\ref{remark:1}. Note that $d^{(n)}{(i, j)}$ is a semi-metric (no triangle inequality).


\vspace{-10pt}
\subsection{Calculating the dissimilarity measure }\label{section:error}
This section reviews the calculation of the dissimilarity measures based on \cite{ourspl}. Let $l_n$ and $q_n$ refer to a sequence of modes representing the ground truth and detected trajectories, respectively, where $l_n=[l_{n-1},i]$, $q_n=[q_{n-1},j]$, and $i,j\in \{1,...,r\}$. Let $\bm{e}_n^{(l_n;q_n)}=\bm{x}_n^{l_n}-{\bm{x}}_{n|n}^{(l_n;q_n)}$ be the conditional estimation error term based on these trajectories, calculated recursively given the trajectories $l_n$ and $q_n$ as 
\begin{align}
    &\bm{e}_n^{(l_n;q_n)}=([\bm{A}^{i}-\bm{A}^j+\bm{K}_n^{q_n}\bm{H}_n\bm{A}^j-\bm{K}_n^{q_n}\bm{H}_n\bm{A}^{i}]\bm{x}_{n-1}^{l_{n-1}}\nonumber\\
    &+(\bm{I}-\bm{K}_n^{q_n}\bm{H}_n)(\bm{A}^j\bm{e}_{n-1}^{(l_{n-1};q_{n-1})}+\bm{\nu}_n) -\bm{K}_n^{q_n}\bm{\omega}_n),
    \label{eq:switch_e_rec}
\end{align}
where $\bm{x}_{n}^{l_{n}}$, ${\bm{x}}_{n|n}^{(l_{n};q_n)}$, $\bm{e}_{n}^{(l_{n};q_{n})}$  refer to the ground truth state variable, the estimated state variable, and the estimation error, respectively, given trajectory $l_n$ occurs and trajectory $q_n=[q_{n-1},j]$ is used for estimation at time $n$. $\bm{K}_n^{q_{n}}$ is the KF gain assuming that trajectory $q_n=[q_{n-1},j]$ has been detected at time $n$. The error $\bm{e}_n=\bm{x}_n-\bm{x}_{n|n}$ may then be written as
\begin{align}
    \bm{e}_n =& \sum_{l_n}\sum_{q_n}\delta_{l_n;q_n}\bm{e}_n^{(l_n;q_n)} , \label{eqn:switche}
\end{align}
where $\delta_{l_n; q_n}$ equals one when $l_n$ occurs and $q_n$ is used in estimation, and 0 otherwise. The mean is calculated as
\begin{align}
    \E[\bm{e}_n]&=\sum_{l_n}\sum_{q_n}\pi_{l_n;q_n}\E[\bm{e}_n^{(l_n;q_n)}],
\end{align}
 where $\pi_{l_n;q_n}$ is the probability that trajectory $l_n$ occurs and trajectory $q_n$ is detected, which may also be calculated recursively. Similarly, we compute $\E[\bm{e}_n^T\bm{e}_n]=\sum_{l_n}\sum_{q_n}\pi_{l_n;q_n}\E[(\bm{e}_n^{(l_n;q_n)})^T{(\bm{e}_n^{(l_n;q_n)})}]$ and the covariance $ \C(\bm{e}_n) =\E[\bm{e}_n\bm{e}_n^T]-\E[\bm{e}_n]\E[\bm{e}_n]^T$. The statistics of the error may be calculated by quantifying and propagating the mismatch error at each time step recursively. Using this formulation, one is able to calculate both directed and undirected dissimilarity measures. Specifically for sets $\mathcal{P},\mathcal{Q}$,
 \begin{align}
     \epsilon_{\mathcal{P}|\mathcal{Q}}^{(n)}=\E[\bm{e}_n^T(\mathcal{P}|\mathcal{Q})\bm{e}_n(\mathcal{P}|\mathcal{Q})]
 \end{align}
  where $\bm{e}_n(\mathcal{P}|\mathcal{Q})=\sum_{l_n \in \mathcal{P}^n}\sum_{q_n \in \mathcal{Q}^n}\delta_{l_n;q_n}\bm{e}_n^{(l_n;q_n)}$, and $\mathcal{P}^n$ refers to all combinations of modes in $\underbrace{\mathcal{P}\times \mathcal{P} \times \dots \times \mathcal{P}}_n$.


\vspace{-12pt}
 \subsection{Clustering}
 \label{sec:clustering}
 
A lower complexity model can be determined for an SLDS by clustering the SLDS modes based on the proposed graph structure. The following theorem quantifies the effect of model complexity reduction on the estimation error in terms of the proposed dissimilarity measure.
\begin{assumption}\label{assumption:1}
The center of each cluster is randomly selected from its members based on their prior occurrence probability.
\end{assumption}

\begin{assumption}\label{assumption:2}
The likelihood of the ground truth mode is the largest (with probability one) compared to the likelihood of all other modes, and the likelihood of the cluster that contains the ground truth mode is the largest (with probability one) compared to the likelihood of all other clusters.
\end{assumption}
\begin{theorem}\label{theorem:1}
Let the {MAP} estimator of the random variable $\bm{x}_n$ that uses the {ground truth} SLDS model including all nodes in the graph $G^*=\{1,...,r\}$ be denoted as $\bm{x}_{n|n}^*$. Let the {MAP} estimator when the ground truth nodes are grouped into $m$ clusters with graph centers $\hat{G}=\{g_1',...,g_m'\}$ be denoted as $\hat{\bm{x}}_{n|n}$.  Given assumptions 1-2,
\begin{align}
\E[(&\bm{x}_n-\hat{\bm{x}}_{n|n})^2]=\E[(\bm{x}_n-\bm{x}_{n|n}^*)^2]\\
&\quad \quad +\sum_{i=1}^m\sum_{j\in \mathcal{S}_i}\sum_{k\in \mathcal{S}_i}\mathds{1}\{j\neq k\} p_i(k)({\pi_k+\pi_j}) d^{(n)}_{j\rightarrow k},\nonumber
  \label{equation:th1}
  \end{align}
  where $\pi_j$ is the prior occurrence probability of each node, $\mathcal{S}_i$ is the set of nodes in cluster $i$, $p_i(k), k\in \mathcal{S}_i$ is the prior probability of node $k$ normalized with respect to the nodes in cluster $i$, and $d^{(n)}_{j\rightarrow k}$ is the directed dissimilarity measure edge.
\end{theorem}

\begin{proof}
\begin{align*}
  &\E[(\bm{x}_n-\hat{\bm{x}}_{n|n})^2-(\bm{x}_n-{\bm{x}}_{n|n}^*)^2]\nonumber\\
  &=\E\big[\E[(\bm{x}_n-\hat{\bm{x}}_{n|n})^2-(\bm{x}_n-{\bm{x}}_{n|n}^*)^2 \mid j\textit{ occurs}]\big]\nonumber\\
  &=\sum_{i=1}^m\sum_{j\in \mathcal{S}_i} \pi_j \E[(\bm{x}_n-\hat{\bm{x}}_{n|n})^2-(\bm{x}_n-{\bm{x}}_{n|n}^*)^2 \mid j \textit{ occurs}]\nonumber\\
    &=\sum_{i=1}^m\sum_{j\in \mathcal{S}_i} \pi_j \E[(\bm{x}_n-\hat{\bm{x}}_{n|n})^2-(\bm{x}_n-{\bm{x}}_{n|n}^*)^2 \mid j \textit{ occurs} ] \nonumber\\
    &\quad\quad\times \mathds{1}\{j =g'_i\}\nonumber \\
    &\quad+\sum_{i=1}^m\sum_{j\in \mathcal{S}_i}\pi_j \E[(\bm{x}_n-\hat{\bm{x}}_{n|n})^2-(\bm{x}_n-{\bm{x}}_{n|n}^*)^2 \mid j \textit{ occurs}]\nonumber\\
    &\quad\quad \times\mathds{1}\{j \neq g'_i\} \nonumber\\
  &\stackrel{(a)}=0 + \sum_{i=1}^m\sum_{j\in \mathcal{S}_i} \pi_j\E[(\bm{x}_n-\hat{\bm{x}}_{n|n})^2-(\bm{x}_n-{\bm{x}}_{n|n}^*)^2 \mid  \nonumber\\
  &\quad\quad j \textit{ occurs}, j\neq g'_i, g'_i \textit{ is detected}] \quad \nonumber\\
  &\stackrel{(b)}=\sum_{i=1}^m\sum_{j\in \mathcal{S}_i} \sum_{k\in \mathcal{S}_i} \pi_jp_i(k)\E[(\bm{x}_n-\hat{\bm{x}}_{n|n})^2-(\bm{x}_n-{\bm{x}}_{n|n}^*)^2 \mid  \nonumber\\
  &\quad\quad  j \textit{ occurs}, g'_i=k \textit{ is detected}] \quad \nonumber\\
  &=\sum_{i=1}^m\sum_{j\in \mathcal{S}_i}\sum_{k\in \mathcal{S}_i}\mathds{1}\{j\neq k\} \pi_jp_i(k) (\epsilon^{(n)}_{\{k\}|\{j\}}-\epsilon^{(n)}_{\{j\}|\{j\}})\nonumber\\
   &=\sum_{i=1}^m\sum_{j\in \mathcal{S}_i}\sum_{k\in \mathcal{S}_i}\mathds{1}\{j\neq k\} p_i(k)({\pi_k+\pi_j}) d^{(n)}_{j\rightarrow k}
  . 
\end{align*}
  Equation $(a)$ holds because if the occurred node $j$ is the center of its cluster, then $\bm{x}_{n|n}^*=\hat{\bm{x}}_{n|n}$ based on the stated assumptions. Equation $(b)$ is true since node $k$ is selected as the center of its cluster $i$ with probability $p_i(k)$.
\end{proof}


In practice, the decision uncertainty in choosing between the modes of a given SLDS is largest when all modes occur with the same probability. To verify this, let $l_n$ be the random variable referring to different $r^n$ trajectories of modes that may have occurred by time $n$. If the probability of all possible trajectories is equal to $\frac{1}{r^n}$, the uncertainty of the random variable $l_n$ is maximized, and therefore, its detection error is maximized\cite{itbook}. Since clustering in the most uncertain case would induce the largest error (compared to when some prior information is available regarding the mode occurrence), characterizing the performance of an SKF for an SLDS with equiprobable modes provides a robust metric for efficient mode selection.
 

\begin{remark}\label{remark:1}
The result of Theorem~\ref{theorem:1} when modes have equal prior probability of occurrence reduces to
\begin{align}
\E[(\bm{x}_n-\hat{\bm{x}}_{n|n})^2]&=\E[(\bm{x}_n-\bm{x}_{n|n}^*)^2]\\
&\quad \quad +\sum_{i=1}^m\sum_{(j,k)\in \mathcal{S}_i\times \mathcal{S}_i}\mathds{1}\{j\neq k\}\frac{2}{r|\mathcal{S}_i|}d^{(n)}(k,j),\nonumber
  \end{align}
  where $\mathcal{S}_i, \mathcal{S}_i\times \mathcal{S}_i$ are the set of nodes and pairs of nodes in the $i^{th}$ cluster, respectively, $|\mathcal{S}_i|$ refers to the number of nodes in cluster $i$, $d^{(n)}(k,j)$ refers to the undirected dissimilarity measure edge between nodes $k$ and $j$, and $r$ is the total number of modes in the system.
\end{remark}
\begin{remark}\label{remark:2}
\textbf{(Clustering algorithm) }According to the result of Remark \ref{remark:1}, graph clustering for mode reduction is performed by solving the following optimization problem,
\begin{align}
    \min_{\{\mathcal{S}_l\}, l=1,\dots,m}{\sum_{i=1}^m\sum_{(j,k)\in \mathcal{S}_i\times \mathcal{S}_i}\mathds{1}\{j\neq k\}\frac{2}{r|\mathcal{S}_i|}d^{(n)}(k,j)}, 
\end{align}
where the optimization is done with respect to graph structures $\{\mathcal{S}_l\},l=1,\dots,m$ to reduce the SLDS from $r$ modes to $m$ modes. 
\end{remark}

In practice, assumptions~\ref{assumption:1}-\ref{assumption:2} are (approximately) satisfied when 1) inter-cluster dissimilarity, i.e., the edges that connect nodes between different clusters, are much larger than intra-cluster dissimilarity, i.e., edges between nodes inside a cluster, and 2) the nodes inside a cluster are well-separated and can be confidently detected from the noisy observations; clustering under these assumptions (approximately) results in the minimum mean squared estimation. A numerical investigation of the clustering algorithm when the assumptions are exactly/approximately satisfied provides valuable insights and is presented in Section~\ref{section:sim}.

\vspace{-10pt}
\section{Computation vs. error trade off }\label{Section:comp}
Consider a $r$-modal SLDS reduced to a $m$-modal SLDS. {For a MAP estimator, the computational saving by clustering is equivalent to running $r-m$ KF prediction steps}. For an MMSE estimator with order $u$ ($u$ is the memory of past trajectories used by the SKF), the obtained computational saving by clustering is equal to running $r^u-m^u$ KFs \cite{skf}. The computational cost of KF (prediction step) for a $z$-dimensional state variable is $O(z^3)$. Therefore, the computational savings obtained for {MAP and} MMSE KF{s} by reducing the number of modes to $m<r$ {are} {$(r-m)\times O(z^3)$,} and $(r^u-m^u)\times O(z^3)$, {respectively}. 
\vspace{-.1cm}
\section{Numerical experiment}\label{section:sim}
Consider a 60-dimensional state space model simulated based on a diffusion process. The diffusivity may switch between $4$ values ($\eta_1 = 0.01$, $\eta_2 = 1$, $\eta_3=0.19$, and $\eta_4 = 0.71$), the measurement operator is $\bm{H}=\bm{I}_{60\times 60}$, and $\bm{Q}_i=0.0004\bm{I}_{60\times 60}$ and $\bm{R} = 0.04\bm{I}_{60\times 60}$. Intuitively, one would expect clustering modes associated with comparable diffusivities to increase the MSE by only a small margin relative to the optimal estimator's MSE, which will be later verified by both the proposed clustering algorithm and Monte Carlo (MC) simulation results. We assume all mode priors and transition probabilities to be equal, and the state variable's initial value $\bm{x}_0$ such that $\E[\bm{x}_0[i]] = 1$ for indices $i=20,\dots,40$, and it is $0$ for all other indices. Also, $\C(\bm{x}_0)=0.01\bm{I}_{60\times 60}$. 

To construct the graph, the undirected dissimilarity matrix for the first time step is calculated as $D = [d^{(1)}(i,j)]_{i,j\in \{1,\dots,r\}}$,
\begin{align*}
D &= 
\begin{bmatrix}
     0&8.86&0.20& 4.28 \\
     8.86& 0&6.17&0.84\\
     0.20& 6.17&0&2.84\\
     4.28&0.84&2.84&0
\end{bmatrix}.
\end{align*}



\begin{figure}[!tbp]
  \centering
   \subfloat[]{\includegraphics[width=0.23\textwidth]{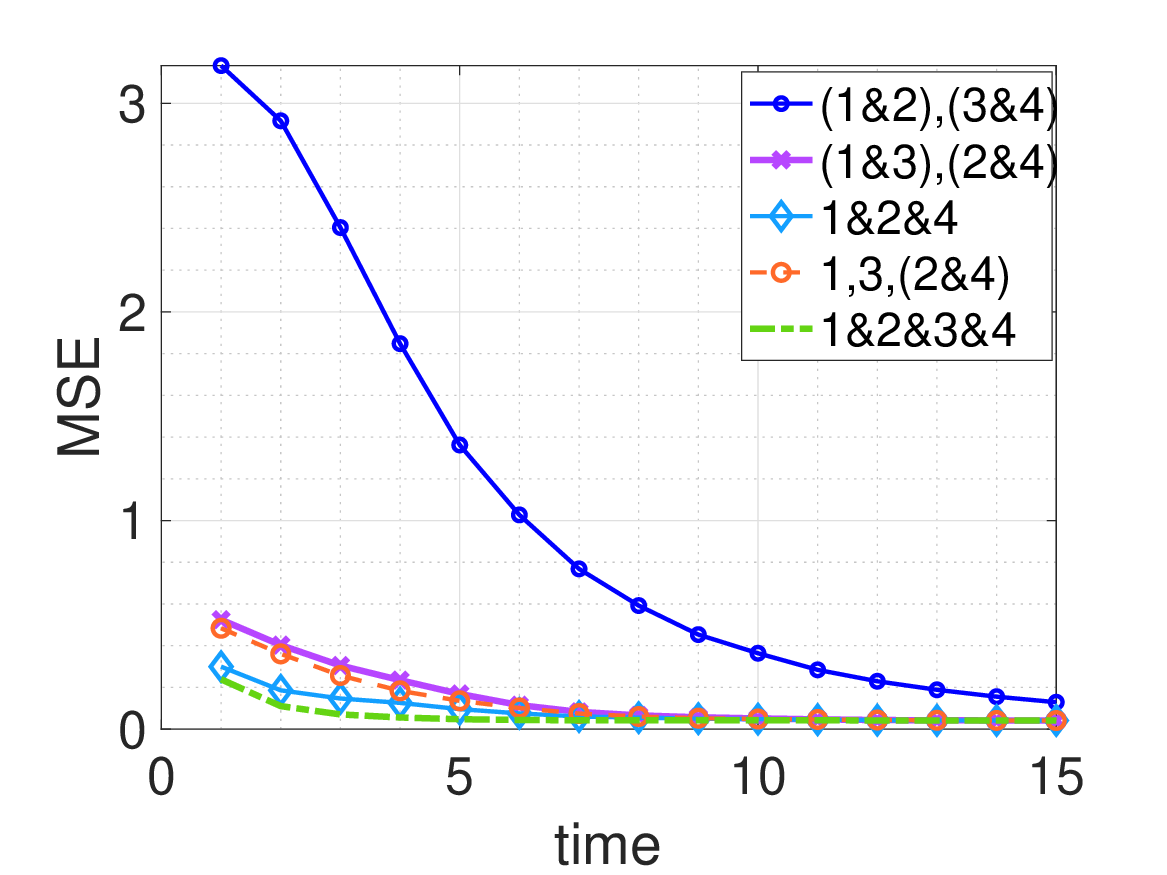}\label{fig:exx1}}
\hfill
  \subfloat[]{\includegraphics[width=0.23\textwidth]{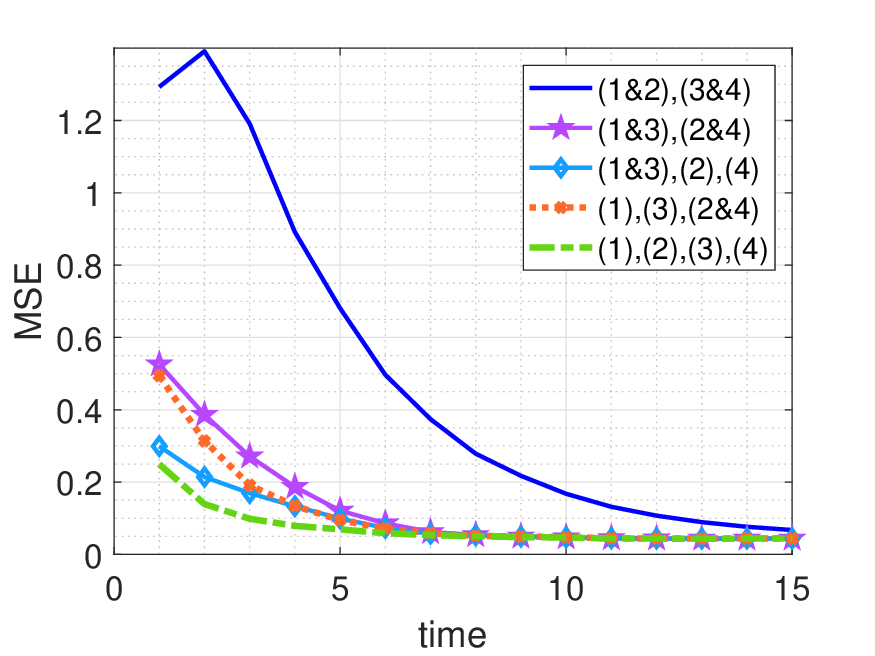}\label{fig:exx2}}
  \caption{  (a) perfect mode detection, (b) mode detection from noisy observations. MC estimation error using a SKF with the ground truth model as well as the reduced models shows that the proposed algorithm can detect the best clusters using the calculated dissimilarity measure matrix in both cases.}
\end{figure}

Let $(a\& b)$ represent a cluster that contains nodes $a$ and $b$. Clustering according to the result of Thm.~\ref{theorem:1} and Remarks ~\ref{remark:1}-\ref{remark:2} shows that three most efficient structures in terms of estimation accuracy are $\{(1\&3),2,4\}$, $\{1,3,(2\&4)\}$, and $\{(1\&3),(2\&4)\}$ (among the set of all possible $1,2,3$-modal structures). For verification, a data set is generated by the 4-modal SLDS for $15$ time steps through $1000$ MC simulations. The estimation error using a 4-modal SKF and complexity reduced SKFs are calculated for two cases, one where the estimator always detects the correct mode (the assumptions of Theorem~\ref{theorem:1} are satisfied), and one where the estimator detects the modes based on the calculated likelihoods of each mode using noisy measurements. The estimation error for both cases are presented in Fig.~\ref{fig:exx1}-\ref{fig:exx2} and show that the decision based on the MC simulations is consistent with the decision made based on the graph dissimilarity matrix. The additional estimation error by clustering the modes compared to the case when using the ground truth SLDS for some example cases including $\{(1\&3),2,4\}, \{(1\&3),(2\&4)\},\{(1\&2), (3\&4)\}$ are computed for the first time step as $0.05, 0.26, 2.92$. As seen in Fig.~\ref{fig:exx2} for the ideal case, the computed additional errors are exact, and this verifies the correctness of the theorem. As observed in Fig.~\ref{fig:exx2} for the realistic case, the calculated additional error is closer to the expected MC-obtained error for the first two mode structures, namely, $\{(1\&3),2,4\}$ and $\{(1\&3),(2\&4)\}$, compared with the last structure $\{(1\&2), (3\&4)\}$. The value obtained for the structure $\{(1\&2), (3\&4)\}$ is not consistent with the MC simulated error since the assumptions mentioned in Theorem~\ref{theorem:1} are not (even approximately) satisfied for this structure. Specifically, node $1$ is more probable to be detected by node $3$, which is not the center of the cluster that contains node $1$.

The run time of {MMSE} SKFs {with order $1$ based on the clustered modes and} using a system with $1.4$ GHz Core i5 processor is presented below:
\begin{center}
 \begin{tabular}{|c |c |c |c |c |c|} 
 \hline
   & $[1,2,3,4]$ &  $[1,3,(2\&4)]$ &$[(1\&3),(2\&4)]$\\ [0.5ex] 
  \hline
  Run time (sec) & 0.13& .09&.05\\ [0.5ex] 
 \hline
 \end{tabular}
 \label{table:1}
\end{center}

Note that the dissimilarity between the modes reduces as the state variables get close to zero. Therefore, the error induced by clustering is at its largest in the first few time steps if all dynamic modes of the SLDS are stable. As a result, calculating the graph structure for each time step is unnecessary and the structure derived for the first time step can be used for optimal estimation at all time steps (this can be verified based on the MC estimates of clustering structures that exactly/approximately satisfy assumptions \ref{assumption:1}-\ref{assumption:2}). Moreover, the optimization of Remark~\ref{remark:2} for the experiment is solved by brute force search over the space of all possible clustering structures. In case the number of modes is large such that brute force search is computationally intractable, one may use greedy algorithms for solving the optimization. 




\section{Summary and Conclusions}\label{section:con}
A computationally tractable graph-based representation along with a clustering approach is developed to reduce the switching mode cardinality of an SLDS and perform computationally efficient estimation. Furthermore, the induced error by clustering is quantified in terms of the mean squared error. The numerical results demonstrate successful performance of the clustering algorithm and verify the quantification of the induced error. The developed clustering algorithm enables computationally efficient state estimation in SLDS that switch between a large number of modes.


\bibliography{main}{}
\bibliographystyle{IEEEtran}










\end{document}